\documentclass[letterpaper,twocolumn,aps,prl,amsmath,amssymb,floatfix]{revtex4}
\usepackage{amsmath,amssymb,amsfonts,mathrsfs}
\usepackage{color}
\usepackage{epsfig}
\usepackage{psfrag}
\usepackage{graphicx}
\usepackage{color}

\begin{document}

\title{Viscosity of a Multi-channel One-Dimensional Fermi Gas}

\author{Wade DeGottardi$^{1,2}$ and K. A. Matveev$^3$}
\affiliation{\small{$^1$ Institute for the Research in Electronics and Applied Physics, University of Maryland, College Park, Maryland 20742, USA}}

\affiliation{\small{$^2$Joint Quantum Institute, NIST/University of Maryland, College Park, Maryland, 20742, USA}}

\affiliation{\small{$^3$ Materials Science Division, Argonne National Laboratory, Argonne, Illinois 60439, USA }}
\date{\today}

\begin{abstract}
Many one-dimensional systems of experimental interest possess multiple bands arising from shallow confining potentials. In this work, we study a gas of weakly interacting fermions and show that the bulk viscosity is dramatically altered by the occupation of more than one band. The reasons for this are two-fold: a multichannel system is more easily displaced from equilibrium and the associated relaxation processes lead to more rapid equilibration than in the single channel case. We estimate the bulk viscosity in terms of the underlying microscopic interactions. The experimental relevance of this physics is discussed in the context of quantum wires and trapped cold atomic gases.
\end{abstract}

\maketitle

The exotic phenomena displayed by one-dimensional (1D) fermion systems---most notably, spin-charge separation and charge fractionalization---indicate that the assumptions of Landau's Fermi liquid theory fail dramatically in one dimension~\cite{deshpande,haldane,giamarchi}. Ultimately, these exotic phenomena arise from the interplay of the Pauli exclusion principle and the restrictions of one-dimensional motion. These aspects also play a crucial role in determining the nature of quasiparticle scattering which in turn controls the relaxation of these systems to equilibrium~\cite{beyond}. For instance, the dual requirements of energy and momentum conservation in one dimension preclude the thermalization of a liquid composed of spinless fermions via two-particle collisions. Instead, relaxation proceeds predominantly through three-particle processes~\cite{andreev}.

Many of the one-dimensional systems of experimental interest are actually \emph{quasi}-one dimensional: transverse modes are accessible at energies lower than the system's bandwidth, thus giving rise to a multichannel system. The multichannel nature of quantum wires is dramatically revealed by the multiple quantized conductance plateaus exhibited in numerous experiments~\cite{vanWees,pepper}. Gases of spinless 1D fermions have been realized in experiments in which clouds of alkali metal atoms are trapped in strongly anisotropic potentials, i.e., $\omega _\| \ll \omega_\perp$, where $\omega_\parallel$ ($\omega_\perp$) is the longitudinal (transverse) confinement frequency of the trap~\cite{rev2}. Occupation of multiple bands is readily controlled and occurs if the number of atoms in the trap exceeds $\omega_\perp / \omega_\|$. Given the experimental prevalence of multichannel systems, we are lead to a natural question. How does the presence of multiple bands alter the macroscopic properties of a 1D system vis-\`{a}-vis strictly single-band systems?

We address this question by evaluating the bulk viscosity of a multichannel system of weakly interacting 1D spinless fermions. In many cases, hydrodynamics captures the behavior of 1D and quasi-1D systems of electrons~\cite{hydro,moll,jong,molenkamp,yu} and cold atoms~\cite{joseph}. Like other hydrodynamical quantities, the bulk viscosity provides a powerful means of connecting microscopic processes to macroscopic responses, thereby linking theory to experimentally relevant quantities~\cite{landau-kinetics}. The bulk viscosity characterizes the dissipation that arises when the fluid's density changes~\cite{landau-fluid}. For the case of transport in a quantum wire, the bulk viscosity controls the extent to which inhomogeneities affect its electrical and thermal conductance~\cite{hydro,transport}. In cold atomic systems, the bulk viscosity is directly related to the damping of breathing modes of the atomic liquid~\cite{schafer,vogt}.

Our central result is that the multichannel nature of the liquid dramatically alters the bulk viscosity. There are two competing aspects at play. First, each component of the gas, i.e., all the particles in a given band, responds to changes in the total fluid density in a slightly different manner. Multichannel systems thus tend to be more easily disturbed from equilibrium; this tends to increase $\zeta$. Second, these systems can relax via two-particle inter-channel collisions. For single channel systems, the relaxation rates are considerably slower given that they necessarily involve three particle collisions which are suppressed at low temperatures. The dramatically increased relaxation rates for multichannel systems has the effect of decreasing $\zeta$. We explore the interplay of these two tendencies and determine the form of $\zeta$ in the multichannel case. Through the identification of the specific two-particle collisions which dominate the relaxation properties, we provide an estimate for the bulk viscosity in terms of the underlying interactions. We discuss the relevance of this physics to quantum wires and gases of laser-trapped cold atoms.

Turning to the specific model of interest, we consider a multichannel system of weakly interacting spinless fermions with dispersion relations
\begin{equation}
\varepsilon_i(p) = \frac{p^2}{2m} + \Delta_i,
\end{equation}
where $\Delta_i$ is the energy to occupy the $p=0$ state of the $i^{\textrm{th}}$ band ($\Delta_i$ is taken to be increasing in $i$). For $\Delta_j < \mu < \Delta_{j+1}$, the system has $j$ occupied bands, with fermions filled to Fermi momenta $\pm k_{F,i}$ in the $i^{\textrm{th}}$ band. The low energy (intra-band) excitations are characterized by Fermi velocities $v_i$, with $v_{i} = \sqrt{2(\mu - \Delta_i)/m}$.

We evaluate the bulk viscosity $\zeta$ in the regime in which the temperature $T \ll |\mu - \Delta_i|$ for all $i$ by considering the dissipation arising from viscous forces. For a fluid with velocity $u(x)$, the rate at which energy is dissipated (per unit length) due to a non-uniform fluid flow is given by
\begin{equation}
w = \zeta \left( \partial_x u \right)^2.
\label{eq:w}
\end{equation}
The quantity $\zeta$ may thus be obtained by evaluating $w$ which can be determined from entropy considerations since $w = T \dot{s}$, where $s$ is the entropy of the system per unit length. Denoting the occupation numbers of a state with momentum $p$ in the $i^{\textrm{th}}$ band by $n_i(p)$, for systems near equilibrium we have that $n_i = f_i + \delta n_i$ where $f_i$ is the equilibrium distribution function
\begin{equation}
f_i(p) = \frac{1}{e^{\left( \varepsilon_i(p) - u p - \mu \right)/T} + 1},
\label{eq:f}
\end{equation}
which reduces to the Fermi-Dirac form for $u = 0$. The quantity $\delta n_i$ is the non-equilibrium correction to the distribution function $f_i(p)$. Now, for fermions $s = - \sum_i \int dp \left[ n_i \ln n_i + \left( 1 - n_i \right) \ln \left(1 - n_i \right) \right]/h$. Differentiating $s$ with respect to time gives $\dot{s} = \sum_i \int dp \ln \left(1/n_i - 1 \right) \dot{n}_i /h$. Expanding this expression to first order in $\delta n_i$ gives
\begin{equation}
w = - T \sum_i \int_{-\infty}^\infty \frac{dp}{h} \frac{\dot{n}_i \delta n_i}{f_i \left( 1 - f_i \right)},
\label{eq:w2}
\end{equation}
since the zeroth order term vanishes.

Equations~(\ref{eq:w}) and (\ref{eq:w2}) demonstrate that $\zeta$ can be obtained if $\dot{n}_i$ and $\delta n_i$ are known. These quantities can be obtained by solving the Boltzmann equation which we write in the form
\begin{eqnarray}
\dot{n}_i(p) &=& I_i\left[ \{ n_j(p) \} \right], \label{eq:ndotI} \\
\dot{n}_i(p) &=& \partial_t n_i(p) + \frac{p}{m} \partial_x n_i(p),
\label{eq:ndotflow}
\end{eqnarray}
where $I$ is the collision integral which describes the relaxation processes. We solve the Boltzmann equation by first expressing $\dot{n}_i(p)$ in terms of $\partial_x u$ using Eq.~(\ref{eq:ndotflow}) and the conservation laws associated with $I$. Then, Eq.~(\ref{eq:ndotI}) will be used to obtain $\delta n_i$.

The leading contribution to $\dot{n}_i(p)$ is obtained by substituting $f_i(p)$ into the right-hand side of Eq.~(\ref{eq:ndotflow}). We consider a specific point in the fluid for which $u = 0$ but $\partial_x u \neq 0$. The continuity equations expressing particle number and energy conservation suggest that both $\mu$ and $T$ will acquire a time dependence at this point. Hence, $f_i(p)$ is given by Eq.~(\ref{eq:f}) for some $u(x)$, $T(t)$ and $\mu(t)$. Substitution of $f_i(p)$ into Eq.~(\ref{eq:ndotflow}) yields $\dot{n}_i(p)$ in terms of $\partial_x u$, $\partial_t T$, and $\partial_t \mu$. These three quantities can be related to each other by two equations that express particle number and energy conservation laws respected by the collisions encoded in $I$. These two equations allow $\partial_t T$ and
$\partial_t \mu$ to be written in terms of $\partial_x u$. In turn, this allows us to write $\dot{n}_i(p)$ in terms of $\partial_x u$ alone, yielding
\begin{equation}
\dot{n}_i(p) =  2  \frac{\partial_x u}{T} f_i \left( 1 - f_i\right) \left( \frac{\sum_j \frac{\Delta_j}{v_j}}{ \sum_j \frac{1}{v_j}} - \Delta_i \right),
\label{eq:ndot3}
\end{equation}
where we have omitted terms proportional to $\varepsilon_i-\mu$ whose contribution to $\zeta$  is subleading at low temperatures.

We now discuss the role of collisions, as described by the collision kernel $I$. Collisions do not alter an equilibrium distribution, i.e., $I_i[\{ f_j(p) \}] = 0$. Hence, it is necessary to include $\delta n_i$. Because the states we consider are near equilibrium, we  linearize $I$ in $\delta n_i$. Furthermore, it is convenient to symmetrize the kernel by introducing $x_i(p)$ where
\begin{equation}
\delta n_i(p) = g_i(p) x_i(p),
\label{eq:symmetrizer}
\end{equation}
and $g_i(p) = \sqrt{f_i(p) \left( 1 - f_i(p) \right)}$. The action of the linearized collision integral on $\{ x_i(p) \}$ can be written in bra-ket notation as
\begin{equation}
| \dot{x} \rangle = - \hat{\Gamma} | x \rangle,
\label{eq:xdot2}
\end{equation}
where the operator $\hat{\Gamma}$ is defined by
\begin{equation}
\hat{\Gamma} | x \rangle = \sum_j \int dp' \, \Gamma_{ij}(p,p') x_j(p').
\end{equation}
Equation~(\ref{eq:symmetrizer}) ensures the symmetry of the kernel, i.e. $\Gamma_{ij}(p,p') = \Gamma_{ji}(p',p)$. Symmetry of the kernel guarantees that the eigenmodes are mutually orthogonal, making it possible to invert $\hat{\Gamma}$~\cite{zeromodes}. This enables us to obtain $\delta n_i$ using Eqs.~(\ref{eq:symmetrizer}) and (\ref{eq:xdot2}). Then, evaluating $w$ using Eq.~(\ref{eq:w2}), we employ Eq.~(\ref{eq:w}) to obtain
\begin{equation}
\zeta = \frac{T}{\left( \partial_x u \right)^2} \langle \dot{x} | \hat{\Gamma}^{-1} | \dot{x} \rangle,
\label{eq:zeta3}
\end{equation}
where the inner product is defined as
\begin{equation}
\langle a | b \rangle = \sum_i \int_{-\infty}^\infty \frac{dp}{h} a_i(p) b_i(p).
\end{equation}
Equation~(\ref{eq:zeta3}) relates $\zeta$ to the relaxation properties of the system.

For any realistic collision process, the inversion of the linearized collision integral is non-trivial. Only in special cases are analytical solutions possible, and in general the solution must be found numerically. However, the essential physics may be extracted by treating $\hat{\Gamma}$ in the relaxation time approximation. In this approximation, the collision integral is taken to be diagonal and every non-zero eigenvalue of $\hat{\Gamma}$ is assumed to be equal to the same constant~\cite{landau-kinetics}. This is tantamount to replacing $\hat{\Gamma}^{-1}$ by $\tau$ in Eq.~(\ref{eq:zeta3}). Noting that $\dot{n}_i(p) = g_i(p) \dot{x}_i(p)$, substitution of Eq.~(\ref{eq:ndot3}) into Eq.~(\ref{eq:zeta3}) gives
\begin{equation}
\zeta = \frac{8 \tau}{h} \sum_i \frac{1}{v_i} \left( \frac{\sum_j \frac{\Delta_j}{v_j}}{ \sum_j \frac{1}{v_j}} - \Delta_i \right)^2.
\label{eq:zeta.relax}
\end{equation}

We observe that in the case of a single-channel system, $\zeta$ in Eq.~(\ref{eq:zeta.relax}) is zero. This illustrates that the multi-channel nature of the system is at the heart of the physics encoded in our expression for $\zeta$. A careful study of the single-channel case for 1D fermions gives non-zero bulk viscosity once the effects of interactions on the excitation spectrum are taken into account~\cite{viscosity}. The simplest non-trivial case for a multi-channel system involves two channels. In this case, Eq.~(\ref{eq:zeta.relax}) can be written in the form
\begin{equation}
\zeta = n m \tau \left( v_{1} - v_{2} \right)^2,
\end{equation}
where $n$ is the total fermion density.

\begin{figure}
\begin{center}
\includegraphics[width = 8cm]{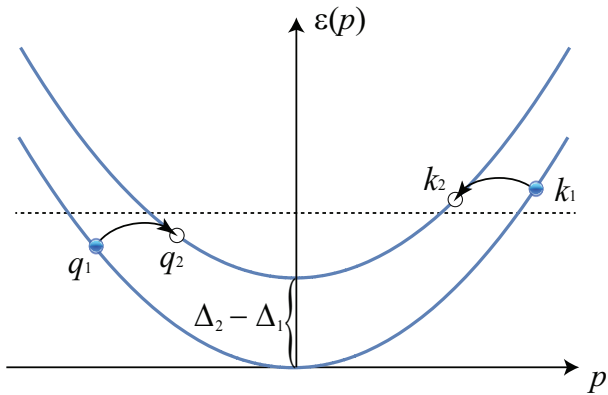}
\label{fig:spectrum}
\caption{The scattering processes which dominate the relaxation properties of a two-channel Fermi gas. In the particular process shown, right- and left-moving fermions in band 1 are both scattered into band 2. The dashed line indicates the chemical potential in the rest frame of the gas.}
\label{fig:2channelscattering}
\end{center}
\end{figure}

The relaxation time $\tau$ may be estimated by Fermi's golden rule once the microscopic processes which allow the system to relax are identified. We begin by considering the two channel case. Because intra-channel processes proceed via three-particle interactions and are strongly suppressed, two-particle inter-channel processes will dominate the relaxation rate.  At low temperatures, the process in which a right- and left-moving particle in band 1 scatter into band 2 (i.e., $1+1 \rightarrow 2+2$), as shown in Fig.~\ref{fig:2channelscattering}, as well as the reverse process represent the most efficient equilibration mechanism. These processes are described by the Hamiltonian
\begin{equation}
H_{\textrm{int}} = \frac{1}{2} \int \! \int \! dx \, dy  \, \psi_1^\dagger(x) \psi_2^\dagger(y) V_{\textrm{eff}}(x-y) \psi_2^{\phantom\dagger}(y) \psi_1^{\phantom\dagger}(x),
\label{eq:rate1}
\end{equation}
where $\psi_i^\dagger(x)$ creates a fermion in band $i$ at position $x$ and $V_{\textrm{eff}}$ is the effective 1D potential between particles in bands 1 and 2. The corresponding scattering rate is given by
\begin{eqnarray}
 \frac{1}{\tau} \! \! &\sim&   \! \! \frac{2\pi}{\hbar} \sum_{q_1 q_2 k_2} \left| \mathcal{M} \right|^2 \delta(\varepsilon_2(k_2) + \varepsilon_2(q_2) - \varepsilon_1(k_1) - \varepsilon_1(q_1)) \nonumber \\
& & \times \delta_{k_2+q_2,k_1+q_1} f_{1}(q_1) \left( 1-f_2(k_2) \right) \left( 1 - f_2(q_2) \right), \label{eq:rate2}
\end{eqnarray}
where $\mathcal{M} = [ \tilde{V}_\textrm{eff}(k_{F1} - k_{F2}) - \tilde{V}_\textrm{eff}(k_{F1} + k_{F2})]/L$ and $L$ is the length of the system. Scaling analysis of this integral reveals that
\begin{equation}
\frac{1}{\tau} \sim \frac{| \mathcal{M} |^2}{\hbar^3 v_1 v_2} \, T.
\label{eq:rate2}
\end{equation}
Thus, combining Eqs.~(\ref{eq:zeta.relax}) and~(\ref{eq:rate2}), we conclude that the bulk viscosity of a two-channel fermionic gas goes as $\zeta \propto T^{-1}$, as $T \rightarrow 0$.

For three or more bands, a variety of inter-channel scattering processes contribute to $\zeta$. However, at low temperatures, only processes of the form $m+m \rightarrow n+n$ contribute to $\zeta$ at leading order in temperature. Given that these processes have rates for which $1/\tau \propto T$, we conclude that the scaling behavior of $\zeta$ also goes as $T^{-1}$ for the case of three or more channels.

In this work, we have studied the viscosity of multichannel 1D fermionic systems. We have seen that the multichannel nature plays a crucial role in determining (a) the form of $\zeta$ and (b) the nature of the relaxation processes. It is natural to think of quasi-1D systems as an interpolation between a 1D system and a Fermi liquid in three dimensions. For both a 1D system and Fermi liquid, $\zeta \propto T^4 \tau$~\cite{viscosity,sykes} and thus is very different from the present case which has $\zeta \propto \tau$. The reason for this difference is that channel number $i$ is analogous to an internal degree of freedom which is known to enhance $\zeta$~\cite{landau-kinetics}. The temperature dependence of $\tau$ also differs in the multi-channel case. For the Fermi liquid, $1/\tau \propto T^2$, giving $\zeta \propto T^2$~\cite{sykes}. For the single channel case in one dimension, three-particle scattering occurs at a rate $1/\tau \propto T^6$ or $T^7$, depending on the nature of the interactions. This gives $\zeta \propto T^{\gamma}$, with $\gamma = -2$ or $-3$ as opposed to $\gamma = -1$ in the multichannel case~\cite{viscosity}.

The divergence of $\zeta$ as $T \rightarrow 0$ underscores its potential relevance to experiment. The relative ease with which multiple band occupation can be controlled both in quantum wires and cold atomic systems makes this physics directly accessible to existing experimental setups. In quantum wires, $\zeta$ plays an important role in the electrical and thermal transport in quantum wires~\cite{hydro,transport}. In such a setting, occupation of multiple channels can be accomplished by tuning a gate voltage, and thus the effects described here are readily investigated in these systems through transport measurements.
The dramatic difference in the viscous properties of single- and multichannel fluids can also be explored in ultra-cold 1D gases of fermions. In this context, $V_{\textrm{eff}}$ in Eq.~(\ref{eq:rate1}) represents $p$-wave scattering which can be tuned by a Feshbach resonance~\cite{granger,gunter}. The most direct probe of $\zeta$ would involve an experiment in which the gas is excited from equilibrium and the dynamics of density oscillations can be observed. In such an experiment, $\zeta$ will be directly related to the decay of the longitudinal breathing mode. Finally, our work sounds a note of caution regarding any experiment endeavoring to study non-equlibrium behavior in 1D quantum systems: the occupation of multiple bands can dramatically affect the relaxation properties and macroscopic responses of the system.

\begin{acknowledgements}
The authors would like to thank A. V. Andreev and B. P. Ruzic for discussions. W.~D. acknowledges support from JQI and IREAP and the Intelligence Community Postdoctoral Research Fellowship Program. Work by K.~A.~M. was supported by the U.S. Department of Energy, Office of Science, Materials Sciences and Engineering Division.
\end{acknowledgements}


\begin{thebibliography}{99}
\bibitem{zeromodes} Inverting $\hat{\Gamma}$ is problematic if this operator has eigenvalues which are equal to zero. Our procedure of using the conservation laws associated with particle number and energy conservation to express $\dot{n}_i$ in terms of $\partial_x u$ in Eq.~(\ref{eq:ndot3}) ensures that the right-hand side of this equation is orthogonal to the corresponding zero modes. The operator $\hat{\Gamma}^{-1}$ in Eq.~(\ref{eq:zeta3}) is thus well-defined since it is restricted to the subspace that excludes zero modes.
\end{thebibliography}
\end{document}